\documentclass[3p, twocolumn]{elsarticle}
\usepackage[boxed,algosection,linesnumbered]{algorithm2e}
\usepackage{epsfig}
\usepackage{bbm}
\usepackage{fullpage}
\usepackage{amssymb}
\usepackage{amsmath}
\usepackage{amscd}
\usepackage{amsthm}
\begin{document}
%
% paper title
% can use linebreaks \\ within to get better formatting as desired
%\title{A Lightweight Simulator For Resource Scheduling Algorithms in a Cloud Data Center Considering Real-time %Multi-dimensional Information}
%\title{A Toolkit For Modeling and Simulation of Resource Scheduling in a Cloud Data Center Considering Real-time %Virtual Machine Allocation}
\title{Open-Source Simulators for Cloud Computing: \\
Comparative Study and Challenging Issues}
%
%
% author names and IEEE memberships
% note positions of commas and nonbreaking spaces ( ~ ) LaTeX will not break
% a structure at a ~ so this keeps an author's name from being broken across
% two lines.
% use \thanks{} to gain access to the first footnote area
% a separate \thanks must be used for each paragraph as LaTeX2e's \thanks
% was not built to handle multiple paragraphs
%
\author[rvt]{Wenhong Tian\corref{cor1}}
\ead{wenhong\_tian@uestc.edu.cn}
\author[rvt]{Minxian Xu}
\ead{xmxyt900@gmail.com}
\author[rvt1]{Aiguo Chen\corref{cor1}}
\ead{agchen@uestc.edu.cn}
\author[rvt]{Guozhong Li}
\ead{aa416517258@hotmail.com}
\author[rvt]{Xinyang Wang}
\ead{uestc\_wxy@qq.com}
\author[rvt]{Yu Chen}
\ead{uestc-chenyu@foxmail.com}

\cortext[cor1]{Corresponding author}
\address[rvt]{School of Information and Software Engineering, University of Electronic Science and Technology of China,China}
\address[rvt1]{School Of Computer Science and Engineering, University of Electronic Science and Technology of China,China}
%\address[focal]{School of Computer Science and Engineering, University of Electronic Science and Technology of China, China}

\begin{abstract}
%\boldmath

Resource scheduling in infrastructure as a service (IaaS) is one of the keys for large-scale Cloud applications. Extensive research on all issues in real environment is extremely difficult because it requires developers to consider network infrastructure and the environment, which may be beyond the control. In addition, the network conditions cannot be controlled or predicted. Performance evaluations of workload models and Cloud provisioning algorithms in a repeatable manner under different configurations are difficult. Therefore, simulators are developed.  To understand and apply better the state-of-the-art of  cloud computing simulators, and to improve them, we study four  known open-source simulators.
%For performance evaluation and modeling of Cloud environments and applications, some simulators are introduced.
They are compared in terms of architecture, modeling elements, simulation process, performance metrics and scalability in performance. Finally, a few challenging issues as future research trends are outlined.
\end{abstract}

\maketitle

% For peer review papers, you can put extra information on the cover
% page as needed:
% \ifCLASSOPTIONpeerreview
% \begin{center} \bfseries EDICS Category: 3-BBND \end{center}
% \fi
%
% For peerreview papers, this IEEEtran command inserts a page break and
% creates the second title. It will be ignored for other modes.
%\IEEEpeerreviewmaketitle
\section{Introduction}
%\newcommand{\printTrueOrFalse}[1]
%{
% \ifthenelse{\equal{#1}{true}}{TRUE}{}
% \ifthenelse{\equal{#1}{false}}{FALSE}{}
%}
%
%\printTrueOrFalse{true}
Cloud computing is developed based on various recent advancements in virtualization, Grid computing, Web computing, utility computing and related technologies. Cloud computing provides both platforms and applications on demand through the Internet or intranet \cite{IEEEhowto:Armbrust}. Some of the key benefits of Cloud computing include the hiding and abstraction of complexity, virtualized resources and efficient use of distributed resources. Some examples of emerging Cloud computing platforms are Google App Engine \cite{IEEEhowto:Google}, IBM blue Cloud \cite{IEEEhowto:IBM}, Amazon EC2 \cite{IEEEhowto:Amazon}, and Microsoft Azure \cite{IEEEhowto:Microsoft}. Cloud computing allows the sharing, allocation and aggregation of software, computational and storage network resources on demand. Cloud computing is still considered in its infancy as there are many challenging issues to be resolved \cite{IEEEhowto:Armbrust}\cite{IEEEhowto:Beloglazov}\cite{IEEEhowto:Buyya2}. Youseff et al. \cite{IEEEhowto:Youseff} establish a detailed ontology of dissecting Cloud into five main layers from top to down: Cloud application (SaaS), Cloud software environment (PaaS), Cloud software infrastructure (IaaS), software kernel and hardware (HaaS), and illustrate their interrelations as well as their inter-dependency on preceding technologies.

Cloud data center can be a distributed network in structure, which is composed of many computing nodes (such as servers), storage nodes, and network devices. Each node is formed by a series of resources such as CPU, memory, network bandwidth and so on. Each resource has its corresponding properties. There are many different types of resources for Cloud providers. The definition and model defined by this paper are aimed to be general enough to be used by a variety of Cloud providers. In this paper, we focus on Infrastructure as a service (IaaS) in Cloud data centers.

In a traditional data center, applications are tied to specific physical servers that are often over-provisioned to deal with workload surges and unexpected failures \cite{IEEEhowto:Singh}. Such configuration rigidity makes data centers expensive to maintain
with wasted energy and floor space, low resource utilizations and significant management overheads.
With virtualization technology, today's Cloud data centers become more flexible, secure and on-demand allocating. %With virtualization,
%Cloud data centers should have ability to migrate an
%application from one set of resources to another in a non-disruptive
%manner. Such agility becomes key in modern cloud
%computing infrastructures that aim to efficiently share and
%manage extremely large data centers.

One key technology plays an important role in Cloud data center is resource scheduling. One of the challenging scheduling problems in Cloud data center is to consider allocation and migration of reconfigurable virtual machines and integrated features of hosting physical machines.

It is extremely difficult to research widely for all these problems in real platforms because the application developers can't control and process network environment. What is more, the network conditions cannot be predicted or controlled.

The research of dynamic and large-scale distributed environment can be achieved by building data center simulation system, which supports visualized modeling and simulation in large-scale applications in cloud infrastructure. Data center simulation system can describe the application workload statement, which includes user information, data center position, the amount of users and data centers, and the amount of resources in each data center. Using this information, data center simulation system generates requests and allocates these requests to virtual machines.

By using data center simulation system, application developers can evaluate suitable strategies such as distributing reasonable data center resources, selecting data center to match special requirements, improving resource utilization and load balancing, reducing total energy-consumptions, reducing costs and so on. We will look at some closely related work firstly.

\subsection {Related Work}
There is quite intensive research conducted for cloud simulators. In this paper, we concentrate on open-source simulators which we can easier access. Dumitrescu and Foster \cite{IEEEhowto:Dumitrescu} introduce GangSim tool for grid scheduling. Buyya et al. introduce GridSim \cite{IEEEhowto:Buyya0} toolkit for modeling and simulation of distributed resource management for grid computing.
Calheiros et al. \cite{IEEEhowto:Buyya} introduce modeling and simulations of Cloud computing environments at application level, a few simple scheduling algorithms such as time-shared and space-shared are discussed and compared.
Sakellari et al. \cite{IEEEhowto:Sakellari} complement a survey of mathematical models, simulation approaches and testbeds in cloud computing, which aims to enable researcher to find suitable modelling approach and simulation implementation.
Ikram et al. \cite{IEEEhowto:Ikram} introduce a novel cloud resource management service model and its simulation-based evaluations are mainly focusing on two applications dynamic service composition.
Nuu et al. \cite{IEEEhowto:Huu} propose a scheme for modeling and experimenting combined smart sleep and power scaling algorithms in energy-aware data center networks.
Guérout et al. \cite{IEEEhowto:Guerout} provide a survey on energy-aware simulation techniques with DVFS (Dynamic Voltage and Frequency Scaling). CloudAnalyst \cite{IEEEhowto:Wickremasinghe} aims to achieve the optimal scheduling among user groups and data centers based on the current configuration.
%\cite{IEEEhowto:Wickremasinghe} introduced three general scheduling algorithms for Cloud computing and some results are provided.
Both CloudSim and CloudAnalyst are based on SimJava \cite{IEEEhowto:Howell} and GridSim \cite{IEEEhowto:Buyya0}, which treat a Cloud data center as a large resource pool and consider application-level workloads. Kliazovich et al. \cite{Kliazovich2010} propose an energy-aware simulation environment named GreenCloud for Cloud datacenters at package level. Nunez et al. \cite{IEEEhowto:Nunez} introduce a new simulator of cloud infrastructure named iCanCloud using C++ and compare the performance with CloudSim. Tian et al. \cite{IEEEhowto:Tian2013-2} propose CloudSched, a novel lightweight simulation tool for VM scheduling with lifecycle in Cloud data centers.

\subsection{Comparative Guideline of Open-Source Cloud Simulators}
Cloud simulators can be divided into various categories according to their features. In this section, we will give a brief comparison with different categories by extending the comparison category in [9]. The open-source simulators are selected because we can study their source codes in details, develop new algorithms and improve them if necessary. The four open source simulators, namely CloudSim, iCanCloud, GreenCloud, CloudSched, are representative of many related simulators because we study the architecture design, modeling elements, simulation process, performance metrics and scalability. These simulators have common features such as in architecture, modeling elements, simulation process as well as their own characteristics such as focusing on different service layers and with different performance metrics.
CloudSim is well known simulator for cloud computing, it can be extended easily but currently it does not consider parallel experiments or lifecycles of VMs. The iCanCloud implements parallel experiments but does not consider energy consumption or VM migration. GreenCloud models detailed energy consumptions for different physical components. CloudSched can model lifecycle of requests, and provide different metrics for load-balance, energy efficiency and utilization etc.
Four open source cloud data centers simulators (CloudSim, GreenCloud, iCanCloud, CloudSched) are compared together in Table 1.
\begin{table*}
\scriptsize
\caption{Comparison Guideline}
\begin{center}
\begin{tabular}{l|l|l|l|l}
\hline Items & CloudSim [24] &GreenCloud [9]& iCanCloud [3] &CloudSched [30]
\\\hline
\hline Platform & any & NS2 & OMNET, MPI & any \\
\hline Programming Language & Java & C++/OTcl &C++ & Java \\
\hline Availability & Open Source &Open Source&Open Source & Open Source \\
\hline Graphical Support & Limited (Via CloudAnalyst) &N&N&Y\\
\hline Physical Models& N &Limited (Via Plug-in)&Y&Y\\
\hline Models for public cloud & N &N&Y&Y\\
\hline Parallel experiments & N&N&Y&N \\
\hline Energy Consumption &Y&Y&N&Y\\
\hline Migration algorithms &Y&N&N&Y\\
\hline Simulation time&Seconds&Tens of minutes&seconds&seconds\\
\hline Memory space&small&large&medium&small\\
\hline
\end{tabular} \\
\end{center}
\end{table*}

\textbf{Platform:} The platform that the simulator based on makes it bind with some specific features. CloudSim and CloudSched are both implemented with Java, so they can be executed on any machine installed JVM. While built based on GridSim and SimJava, CloudSim is heavy to execute.
%MDCSim is written in CSIM, a commercial tool that is not open for downloading now.
GreenCloud is an extension of NS2 network simulator, and it's a packet level simulator. As for iCanCloud, it's based on OMNET, which can simulate in-depth physical layer entities.

\textbf{Language:} The languages implemented the simulators are related to the platforms. CloudSim and CloudSched are implemented with Java, while GreenCloud needs combining C++ and OTcl, iCanCloud is in C++.

\textbf{Availability:} The four simulators under discussion are free or open-source, available for public download.

\textbf{Graphical support:} The original CloudSim supports no graphical interface, the graphical interface is supported in CloudAnalyst. However, full support is not provided in CloudAnalyst, only the configurations and results can be presented. So we label it as “limited”, the same reason is also applicable for GreenCloud. CloudSched and iCanCloud support whole scheduling process to be showed on the interfaces.

\textbf{Physical server models:} The details about the simulated components can reflect the precision of the simulator and the validity of the results. iCanCloud and CloudSched provide detailed simulation for physical analogs for the scheduling, which can trace resource utilization in physical servers and rejected requests information. GreenCloud needs to use a plug-in to simulate and then it can even capture the packet loss. CloudSim treats resource pool as a whole.

\textbf{Models for public cloud providers:} Amazon, as a cloud provider, has proposed its VM models and informed that by using these specifications, better scheduling effects could be obtained. Both iCanCloud and CloudSched use the model suggested by Amazon, in which physical machine and virtual machine specifications are pre-defined.

\textbf{Parallel experiments:} Parallel experiments could combine more than one machine to work together to process the tasks. Supporting for multiple machines running experiments together is a main feature of iCanCloud and that feature is not presented in other three simulators.

\textbf{Energy consumption model:} The energy consumption model can enable the simulators to compare energy efficiency of different scheduling strategies and algorithms. Except for iCanCloud, other three simulators can support energy consumption modeling. The energy consumption model implemented in GreenCloud can trace every element in a data center. DVFS energy consumption model is proposed in CloudSim with extension tools. CloudSched provides energy consumption metrics for different scheduling algorithms.

\textbf{Migration algorithms:} Migration algorithms are proposed to satisfy specific objectives, for instance dealing with the overloaded scenario in load balancing applications, reducing the total number of running machines to save total energy consumption, improving the resource utilization and so on. CloudSim and CloudSched support migration algorithms, while other two simulators do not.

\textbf{Scalability:} This mainly means how fast the simulator can run (simulation time) and how much memory space the simulator will consume as the total number of requests is increasing, especially to a large amount. We will provide comparison in performance evaluation.

In summary, CloudSim, GreenCloud, iCanCloud and CloudSched are open source and available to download. CloudSim and GreenCloud offer no graphical interface support; CloudSched and iCanCloud all provide user interface to operate. CloudSched and iCanCloud support physical server models, and GreenCloud supports physical models with a plug-in. In addition, CloudSched and iCanCloud offer models for public cloud providers. Parallel experiments are supported only in iCanCloud, but only iCanCloud does not support energy consumption model. CloudSim and CloudSched implement migration algorithms while others not.
In the following sections, we will provide in-depth comparative study in terms of architecture design, simulation process, elements, performance metrics and scalability in performance.

The organization of remaining parts of this paper is as follows: from section 2 to section 6, detailed comparisons from different views about CloudSim, GreenCloud, iCanCloud and CloudSched are given. Section 2 compares the architecture and main features of these simulators; section 3 compares the way how elements are modeled in different simulators; section 4 presents the basic simulation process and compares minor differences in those simulators; section 5 lists the metrics in use; section 6 shows how performance are evaluated in those simulators; finally conclusions about cloud simulators are given.

\section{Comparison 1: Architecture and Main Features}
In this section, we will discuss the simulators architectures.

%\begin{figure*} [htp!]
%\begin{center}
%%\hfill
%{\includegraphics [width=0.45\textwidth,angle=-0] {CloudSimArchi.pdf}}
%%\hspace*{\fill}
%%\hfill \includegraphics [width=0.45\textwidth,angle=-0]%{3classesChangeExample2.ps}\hspace*{\fill}
%\caption{The Architecture of CloudSim \cite{IEEEhowto:Buyya}}
%\end{center}
%\end{figure*}

\begin{figure} [htp!]
\begin{center}
{\includegraphics [width=0.5\textwidth,angle=-0] {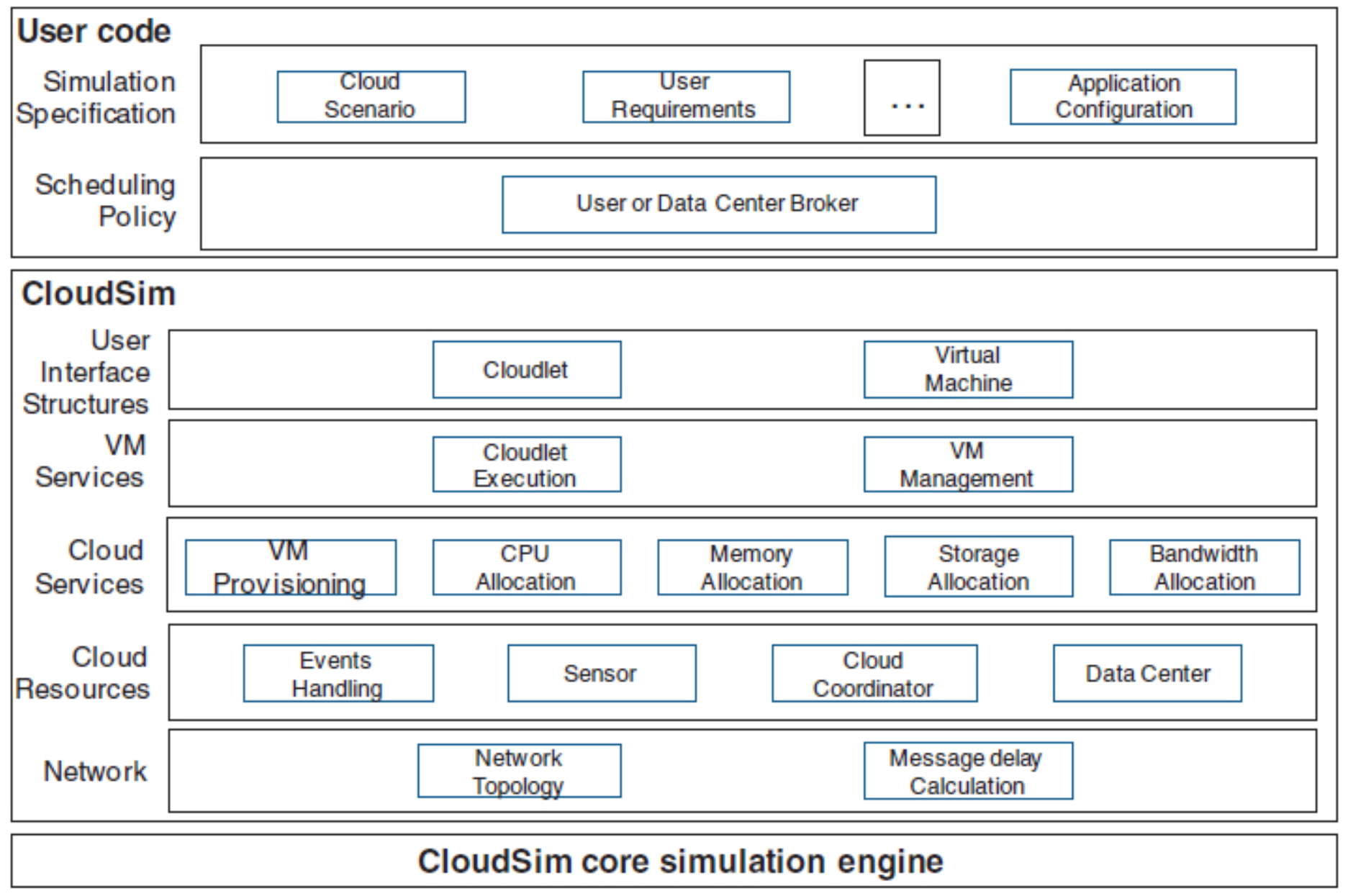}}
\caption{The Architecture of CloudSim \cite{IEEEhowto:Buyya}}
\end{center}
\end{figure}

Fig. 1 shows the multi-layered design and implementation of CloudSim. At the fundamental layer, management of applications, hosts of VMs, and dynamic system states are provided. By extending the core VM provisioning functionality, the Cloud provider can also study the efficiency of different strategies at this layer. As for the top layer, the User Code represents the basic entities for hosts, and through extending entities at this layer, developer can enable the application to generate requests in a variety of approaches and configurations, model cloud scenarios, implement custom applications, and etc.
%\begin{figure*} [htp!]
%\begin{center}
%{\includegraphics [width=0.45\textwidth,angle=-0] {GreenCloudArchi.pdf}}
%]{3classesChangeExample2.ps}\hspace*{\fill}
%\caption{Three-tier data center architecture of GreenCloud\cite{Kliazovich2010}}
%\end{center}
%\end{figure*}

\begin{figure} [htp!]
\begin{center}
{\includegraphics [width=0.5\textwidth,angle=-0] {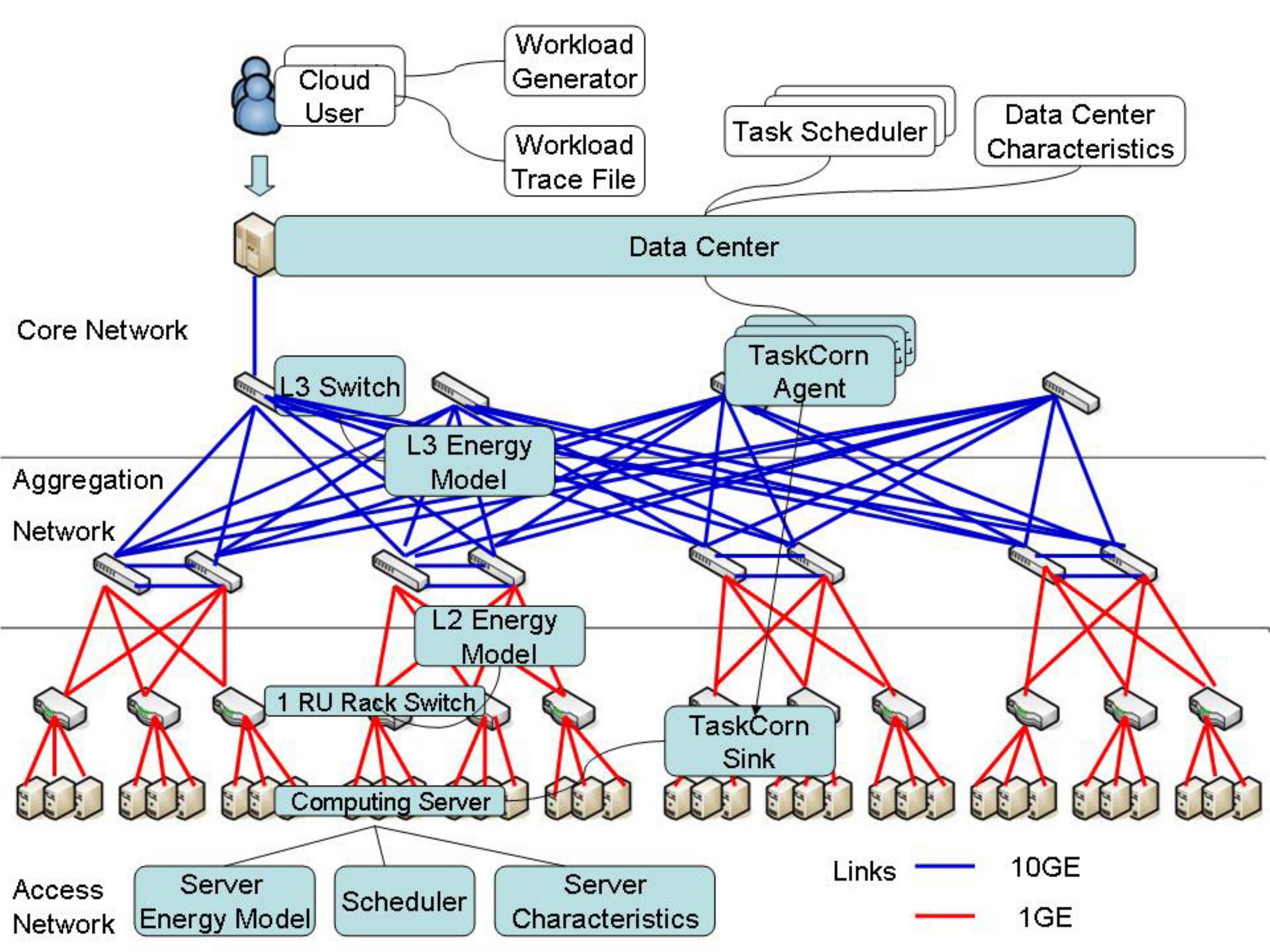}}
\caption{Three-tier data center architecture of GreenCloud\cite{Kliazovich2010}}
\end{center}
\end{figure}

GreenCloud structure could be mapped on the three-tier data center architectures as in Fig. 2, which are the most common architectures. Basically, the architectures are composed of access layers, aggregation layers and cores layers. Servers are placed at the access layer and responsible for task execution. Switches and Links form the interconnection fabric that delivers workload to any of the computing servers for execution at the aggregation layer. The core layer constitutes the workloads that can model various cloud user services.

%\begin{figure*} [htp!]
%\begin{center}
%{\includegraphics [width=0.45\textwidth,angle=-0] {iCanCloudArchi.pdf}}
%{3classesChangeExample2.ps}\hspace*{\fill}
%\caption{The Architecture of iCanCloud \cite{IEEEhowto:Nunez}}
%\end{center}
%\end{figure*}

\begin{figure} [htp!]
\begin{center}
{\includegraphics [width=0.5\textwidth,angle=-0] {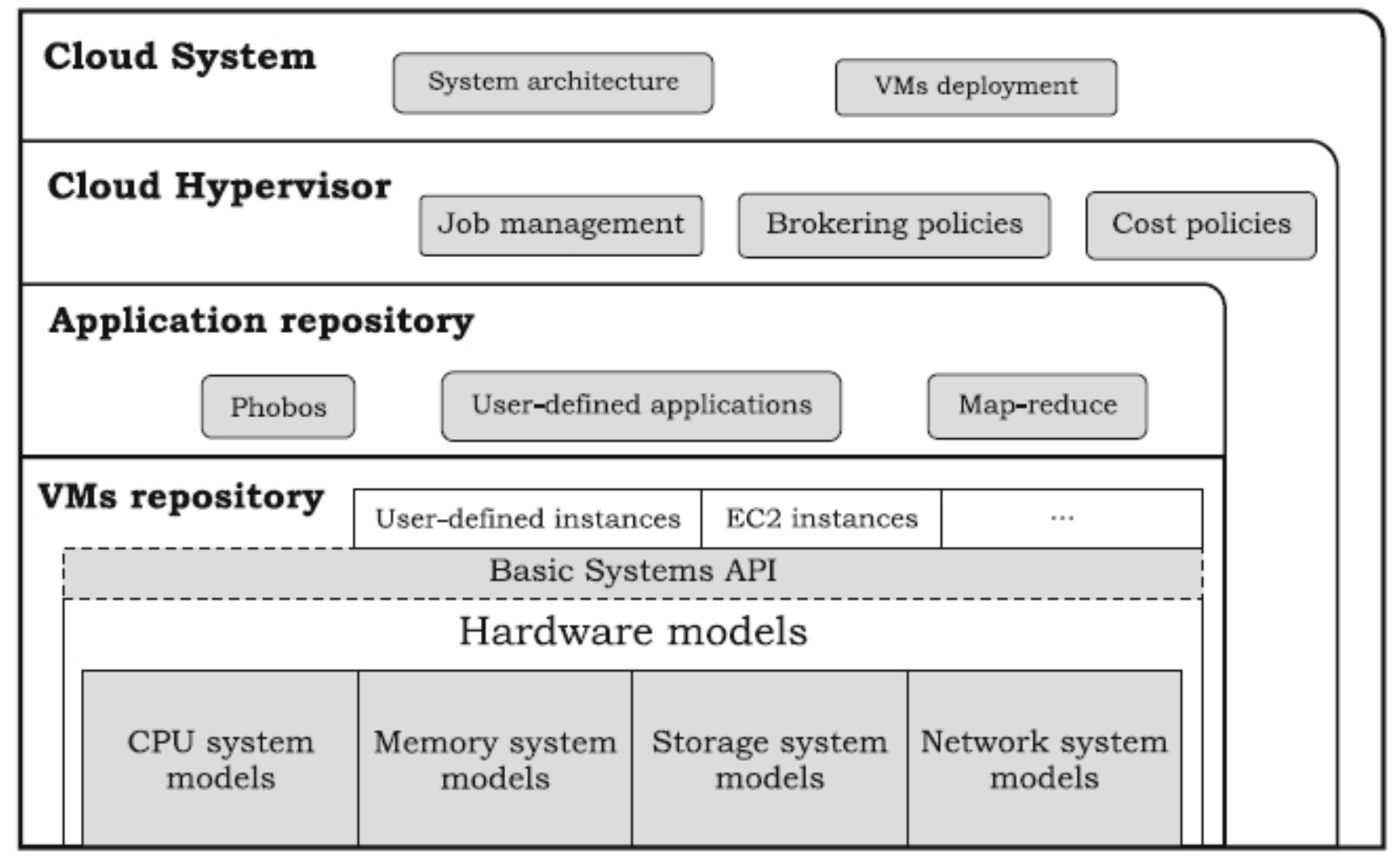}}
\caption{The Architecture of iCanCloud \cite{IEEEhowto:Nunez}}
\end{center}
\end{figure}

The iCanCloud adopts the architecture shown in Fig. 3, which is also a layered architecture. The bottom of the architecture consists of the hardware models layer, which basically contains the models that are in charge of modeling the hardware parts of a system. A set of system calls are connected with the hardware models layer in the basic system’s API module. In this module, a set of system calls are provided as Application Programming Interface for all applications run in a VM. The upper layer is a VMs repository, which contains a collection of VMs previously defined by the user. The cloud hypervisor is at the upper layer that is managing all produced jobs and the instances of VMs where those jobs are executed. As for the top of architecture, it contains a definition of the entire cloud system.

\begin{figure} [htp!]
\begin{center}
{\includegraphics [width=0.45\textwidth,angle=-0] {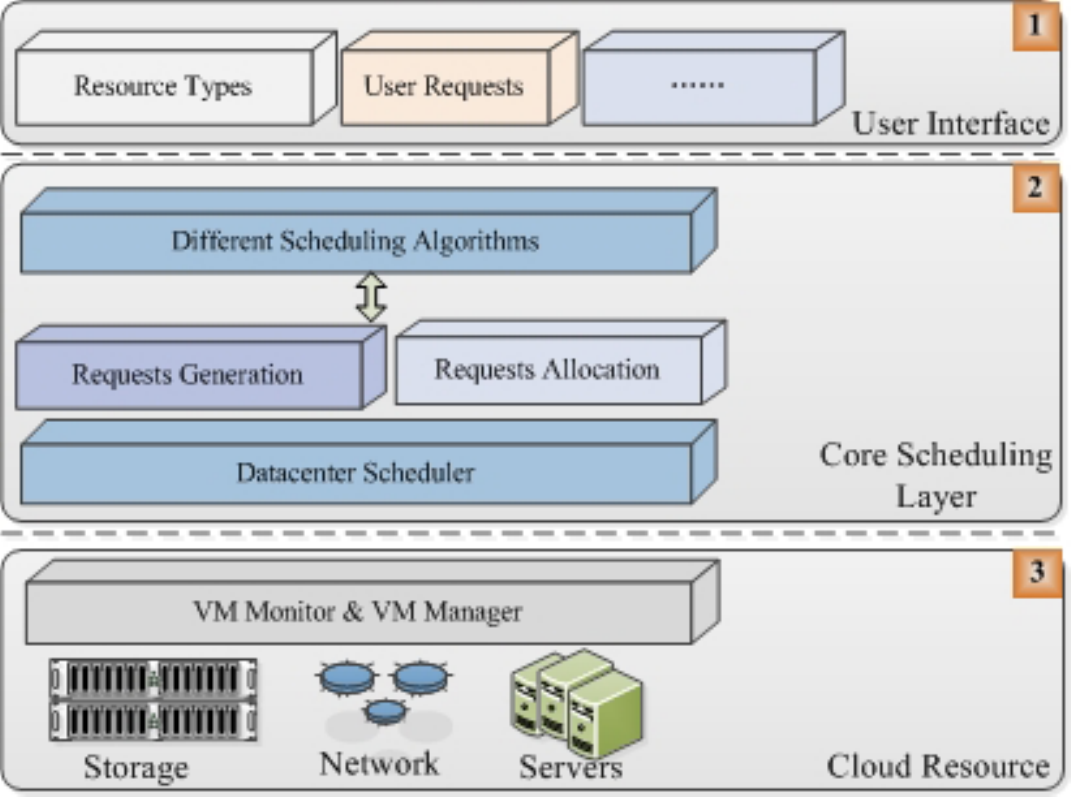}}
\caption{A Simplified Layered Architecture of CloudSched \cite{IEEEhowto:Tian2013-2}}
\end{center}
\end{figure}

CloudSched is implemented under a simplified layered architecture as shown in Fig. 4. From top to bottom layer, at the top layer, there is an interface for a user to select resources and send requests, basically, a few types of virtual machines are preconfigured for a user to choose. The lower layer is the core layer of scheduling: once user requests are generated, those requests are forwarded to next level, which is responsible to choose appropriate data centers and physical machines based on user requests. CloudSched provides support for modeling and simulation of Cloud data centers, especially allocating virtual machines (consisting of CPU, memory, storage and bandwidth etc.) to suitable physical machines. This layer can manage a large scale of Cloud data centers consisting of thousands of physical machines. Different scheduling algorithms can be applied in different data centers based on customers' characteristics. At the bottom layer, there are Cloud resources which include physical machines and virtual machines, both of them consisting of certain amount of CPU, memory, storage and bandwidth etc.
In summary, from the architecture view, the compared simulators all adopt the layered architecture and the layers can be mainly divided into three parts. Each layer is responsible for some basic functions. At the bottom layer, these simulators provide management for servers (in both GreenCloud and iCanCloud) or hosts of VMs (in CloudSim and CloudSched). The upper layer are in charge of scheduling the tasks (comparing efficiency of different algorithms or strategies). At the top layer, interface for users are offered, including configurations or scenarios that can be set by the users in all these simulators. Besides these basic functions, some extra functions are extended in different simulators, while the basic ones are quite similar.

\section{Comparison 2: Building Blocks in Simulators}
In this section, we discuss the building blocks, i.e., the elements modeled in each simulator.

\subsection{Modeling Cloud Data Centers}
In CloudSim and CloudAnalyst, the infrastructure-level services related to the clouds are simulated by modeling the data center entity. In CloudSim, an entity represents an instance of a component, like data center or host. The data center entity manages a number of host entities and these hosts can be assigned to one or more VMs based on allocation policy. Host represents a physical computing server in a Cloud, with processing capability, including CPU, memory, storage, etc. In data center, both hosts and VMs can be managed during their life cycles.

In GreenCloud, elements are modeled based on the multi-tier data center architecture. Servers, switches and links, and workloads constitute the basic elements of GreenCloud. Servers are responsible for task execution, quite similar to the servers in the CloudSim, and workloads can be viewed as the VM requests (tasks) in the CloudSim simulator. As for the switches and links, they form the interconnection fabric that delivers workload to any of the computing servers for execution in a timely manner. The VMs are in a variety of specification in CloudSim or CloudSched, while workloads in GreenCloud are divided into three types: Computational Intensive Workloads, Data Intensive Workloads and Balanced Workloads.

In iCanCloud, the elements model has some differences. The main difference lies in the servers modeling. In iCanCloud, hardware model represents the resources provided in the simulator and VM instances take the place of servers in other simulators. A data center represents a set of Virtual machines, and the VMs are responsible for executing the scheduled jobs, which are a list of tasks submitted by users.

In CloudSched, the core hardware infrastructure related to the Cloud is modeled with a data center component for handling VM requests. The data center component is mainly composed of a set of hosts, which are responsible for managing VMs activity during their life cycles. A host is a component that represents a physical computing node in a Cloud: it is assigned a pre-configured processing capability (expressed in computing power in CPU units), memory, bandwidth, storage, and a scheduling policy for allocating processor cores to virtual machines. A VM could be represented in a similar way like the host.
%Allocation of application-specific VMs to Hosts in a Cloud-based data center is the responsibility of the Virtual Machine Provision component. This component exposes a number of custom methods for researchers, which aids in implementation of new VM provisioning policies based on optimization goals (user centric, system centric). The default policy implemented by the VM Provision is a straightforward policy that allocates a VM to the Host in First-Come-First-Serve (FCFS) basis. The system parameters such as the required number of processing cores, memory and storage as requested by the Cloud user form the basis for such mappings. Other complicated policies can be written by the researchers based on the infrastructure and application demands.

\subsection{Modeling Virtual Machine Allocation}
VM allocation is the process of generating VM instances on hosts that match the critical resources, configurations, and requirements of the Cloud provider. With virtualization technologies, Cloud computing provides flexibility in resource allocation. For example, a PM(Physical Machine) with two processing cores can host two or more VMs on each core concurrently. Only if the total used amount of processing power by all VMs on a host is not more than available capacity in that host, VMs can be allocated.

Taking the widely used example of Amazon EC2 \cite{IEEEhowto:Amazon}, we show that a uniform view of different types of VMs is possible. Table 2 shows eight types of virtual machines from Amazon EC2 online information. The speed per CPU core is measured in EC2 Compute Units, being each C.U. equivalent to a 1.0-1.2 GHz 2007 Opteron or 2007 Xeon processor. We can therefore form three types of different PMs (or PM pools) based on compute units. In real Cloud data center, for example, a physical machine with 2$\times$68.4GB memory, 16 cores$\times$3.25 units, 2$\times$1690GB storage can be provided. In this or similar way, a uniform view of different types of virtual machines is possibly formed. This kind of classification provides a uniform view of virtualized resources for heterogeneous virtualization platforms e.g., Xen, KVM, VMWare, etc., and brings great benefits for virtual machine management and allocation. Customers only need selecting suitable types of VMs based on their requirements. There are eight types of VMs in EC2 as shown in Table 2, where MEM is for memory with unit GB, CPU is normalized to unit (each CPU unit is equal to 1Ghz 2007 Intel Pentium processor \cite{IEEEhowto:Amazon}). Three types of PMs are considered for heterogeneous case as shown in Table 3.
\begin{table}
\footnotesize
\caption{8 types of virtual machines (VMs) in Amazon EC2
}
\begin{center}
\begin{tabular}{|l|l|l||l|}
\hline MEM (GB) & CPU (units)& BW(G)&VM
\\\hline
\hline 1.7 & 1 (1 cores x 1 units)& 160& 1-1(1) \\
\hline 7.5 &4 (2 cores x 2 units) & 850& 1-2(2) \\
\hline 15.0& 8 (4 cores x 2 units) &1690&1-3(3) \\
\hline 17.1& 6.5 (2 cores x 3.25 units) &420&2-1(4)\\
\hline 34.2& 13 (4 cores x 3.25 units) &850&2-2(5)\\
\hline 68.4& 26 (8 cores x 3.25 units) &1690&2-3(6)\\
\hline 1.7& 5 (2 cores x 2.5 units) &350&3-1(7) \\
\hline 7.0& 20 (8 cores x 2.5 units) &1690&3-2(8)\\
\hline
\end{tabular} \\
\end{center}
\end{table}

\begin{table}
\scriptsize
\caption{3 types of physical machines (PMs) in Amazon EC2
}
\begin{center}
\begin{tabular}{|l|l|l|l|l|}
\hline CPU (units) & MEM(G)& BW(G)&$P_{min}$&$P_{max}$
\\\hline
\hline 16(4 cores x 4 units) & 30& 3380G &210W&300W \\
\hline 52(16 cores x 3.25 units) & 136.8&3380G&420W&600W \\
\hline 40(16 cores x 2.5 units) &14&3380G&350W&500W \\
\hline
\end{tabular} \\
\end{center}
\end{table}
CloudSim supports the development of custom application service models that can be deployed within a VM and its users are required to extend the core Cloudlet object for implanting their application services. To be exactly, VMs or jobs in CloudSim, iCanCloud can only be allocated to hosts that have enough resources, like memory, storage, etc.

Workloads in GreenCloud need a complete satisfaction of its two main requirements: computing and communicational, which define the amount of computing that has to be executed before a given deadline and the size of data transfers that must be performed prior, during, and after the workload execution.

Currently CloudSched implements dynamic load-balancing scheduling algorithms, utilization maximization and energy-efficient scheduling algorithms. Other algorithms such as reliability-oriented and cost-oriented etc. can be applied as well.

\subsection{Modeling Customer Requirements}
CloudSim models the customer requirements by deploying VM instances and users can extend the core Cloudlet object for implementing their application services. The VM instance may require some resource such as memory, storage and bandwidth on the host to enable its allocation, which means assign specific cores of CPU, amount of memory and bandwidth to specific VMs.

GreenCloud models customer requirements by configuring the workload arrival rate/pattern to the data center following a predefined distribution (like exponential distribution), or generating requests from traces log files. In addition, different random distributions can also be configured to trigger the time of a workload arrival as well as specify the size of the workload. This flexibility enables users to adopt various choices to investigate network conditions, traffic load, and influences on different switching components. Moreover, the trace-driven workload generation makes it more realistic to simulate the workload arrival process.

In iCanCloud, VMs are the building blocks for creating cloud systems. Both in the application repository and VMs repository, collections of pre-defined models can be customized by user. Those models will be used in order to configure the corresponding jobs that will be executed in a specific instance of a VM in the system. Also, new application models can be easily added to the system.

CloudSched models customer requirements by randomly generating different types of VMs and allocates VMs based on appropriate scheduling algorithms in different data centers. The arrival process, service time distribution and required capacity distribution of requests can be generated according to random processes. The arrival rate of customers' requests can be controlled. Distribution of different types of VM requirements can be set too. A real-time VM request can be represented in an interval vector: vmID(VM typeID, start-time, end-time, requested capacity). For example, vm1(1, 0, 6, 0.25) shows that the request ID is 1, virtual machine is of type 1 (corresponding to integer 1), start-time is 0 and end-time is 6 (here 6 means the end-time is the sixth slot). Other requests can be represented in similar ways. Fig. 5 shows the life cycles of virtual machine allocation in a slotted time window using two PMs, where PM$\#$1 hosts vm1, vm2 and vm3 while PM$\#$2 hosts vm4, vm5 and vm6. Notice that at any slot, the total capacity constraint of a PM has to be met by all VMs allocated on it, and each VM has a start-time, end-time constraint.

In summary, in order to satisfy the flexibility and extendibility of customer requirements, these simulators all provide predefined configurations as well as interfaces for extending. CloudSim can extend the core Cloudlet object; GreenCloud can generate customer requests in trace log file; iCanCloud can modify the application model in the application and VMs repository; CloudSched can change the VM and PM specification in the configuration files.
\begin{figure} [htp!]
\begin{center}
{\includegraphics [width=0.5\textwidth,angle=-0] {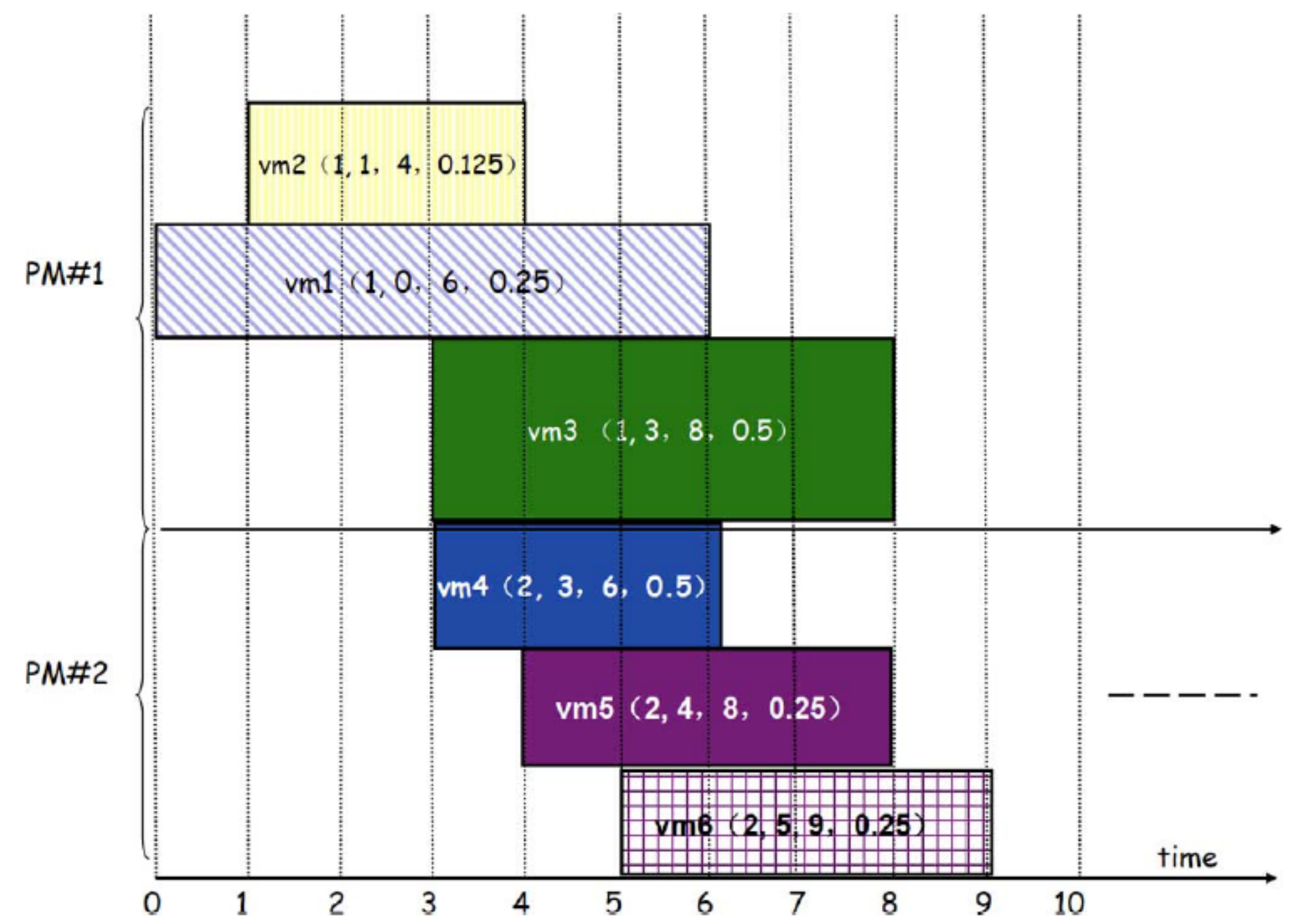}}
\caption{An Example of User Requests and Allocation}
\end{center}
\end{figure}

\section {Comparison 3: Simulation Process}
Generally, the simulation process for cloud data centers can be mainly divided into four parts: 1) generating customer requests; 2) initiating data centers; 3) defining allocation policy; 4) collecting and outputing results. The simulators that we discussed in this paper all have these four parts, though some differences existed when extending the basic parts.

\textbf{Generating customer requests:}
Requests are generated in this phase and prepared to be allocated. In different simulators, the requests generation approaches may vary and preparation process before requests allocation would also have minor differences. Requests in CloudSim, CloudSched are generated as VM instances and put into different queues in different phases, like waiting queue represents the requests are waiting to be executed. Workloads are produced in GreenCloud with its size satisfying exponential distribution. Jobs in iCanCloud can be submitted by user or pre-defined model as list and then be added into the waiting queue to be executed.

\textbf{Initiating data centers:}
In this phase, data center are started to provide resources. The discussed simulators are almost similar in initializing cloud data centers and they initialize the servers/hosts to offer resource like CPU, memory, storage and etc. To be noticed, the servers/hosts may be geographical separated, which means located in different data centers.

\textbf{Defining allocation policies:}
Allocation policy describes scheduling process, including when and how to allocate the specific request to the specific server/host. Allocation policy has a tight relationship with the goal of scheduling. For instance, load balancing and energy saving may use different allocation policies. In CloudSim and iCanCloud, First Come First Service (FCFS) policy is implemented as a basic choice. CloudSched develops some load balancing policies to compare performance and GreenCloud contains DVFS (Dynamic
Voltage Frequency Scaling) policies to evaluate energy saving effects.

\textbf{Collecting and outputting results:}
After the scheduling process is completed, results would be gathered to evaluate the performance of a policy. Except CloudSim, other simulators would present part of simulation results in the user interface. Similarly, with different scheduling goals, evaluated indices would vary. The comparison indices and typical outputs would be introduced in the following sections.

\section{Comparison 4: Performance Metrics}
For different objectives of scheduling, there are different performance metrics. In this section, we discuss some usual metrics that adopted in cloud simulators, like for utilization maximization, load-balancing, energy-efficient goals. Other metrics for different objectives can be extended easily based on these usual metrics. Note that the four simulator use quite different metrics, here we just try to cover the metrics which are applied in the four simulators. Table 4 summaries the metrics name, metrics objective and the simulators that adopt the corresponding metric.
\subsection{Metric for Maximizing Resource Utilization}
In the following, we firstly review two metrics for maximizing resource utilization and these two metrics are the basis for load balancing and energy efficient in the following subsections.\\
\noindent (1). Average resource utilization. Average utilization of CPU, memory, hard disk and network bandwidth can be computed and an integration utilization of all these resources can be used too. \\
(2). The total number of PMs used. It is closely related to the average and whole utilization of a Cloud data center.\\

\subsection{Metrics for Multi-dimensional Load-Balancing}

In view of advantages and disadvantages of existing metrics for resource scheduling \cite{IEEEhowto:Singh}\cite{IEEEhowto:Wood}\cite{IEEEhowto:Zheng}\cite{IEEEhowto:Tian2}, integrated measurement on total imbalance level of Cloud data center and each server are developed for load-balancing strategy [33]. The following parameters are considered:\\
(1). average CPU utilization ($CPU_i^{U}$) of a single CPU $i$: For example, if the observed period is one minute and CPU utilization is recorded every 10 seconds, then $CPU_i^{u}$ is the average of six recorded values of CPU $i$. This metric could represent the average load on a single CPU during a period of observed time.\\
(2). average utilization of all CPUs in a Cloud datacenter: Let $CPU_i^{n}$ be the total number of CPUs of server $i$, then the average utilization of all CPUs on server $i$ is
\begin{equation}
CPU_u^{A}=\frac{\sum_{i}^{N}CPU_i^{U}CPU_i^{n}}{\sum_{i}^{N}CPU_i^{n}}
\end{equation}
where $N$ is the total number of physical servers in a Cloud datacenter. Similarly, average utilization of memory, network bandwidth of server $i$, all memories and all network bandwidth in a Cloud datacenter can be defined as $MEM_i^{U}$, $NET_i^{U}$, $MEM_u^{A}$, $NET_u^{A}$ respectively.\\
(3). integrated load imbalance value ($ILB_i$) of server $i$: Variance is widely used as a measure of how far a set of numbers is spread out from each other in statistics. Using variance, an integrated load imbalance value ($ILB_i$) of server $i$ is defined as:
\begin{equation}
\footnotesize
\frac{(Avg_i-CPU_u^{A})^2+(Avg_i-MEM_u^{A})^2+(Avg_i-NET_u^{A})^2}{3}
\end{equation}
where
\begin{equation}
\small
Avg_i=(CPU_i^{U}+MEM_i^{U}+NET_i^{U})/3
\end{equation}
$ILB_i$ could be applied to indicate load imbalance level comparing utilization of CPU, memory and network bandwidth of a single server itself.\\
(4). the imbalance value of all CPUs, memories and network bandwidth:
Using variance, the imbalance value of all CPUs in a data center is defined as
\begin{equation}
IBL_{CPU}=\sum_{i}^{N}(CPU_i^{U}-CPU_u^{A})^2
\end{equation}
Similarly, imbalance values of memory ($IBL_{mem}$) and network bandwidth ($IBL_{net}$) can be calculated.
Then total imbalance values of all servers in a Cloud datacenter is given by
\begin{equation}
IBL_{tot}=\sum_{i}^{N}ILB_i
\end{equation}\\
(5). average imbalance value of a physical server $i$:
The average imbalance value of a physical server $i$ is defined as
\begin{equation}
IBL_{avg}^{PM}=\frac{IBL_{tot}}{N}
\end{equation}
where $N$ is the total number of servers. As its name suggests, this value can be used to measure average imbalance level of all physical servers.\\
(6). average imbalance value of a Cloud datacenter (CDC):
The average imbalance value of a Cloud datacenter (CDC) is defined as
\begin{equation}
IBL_{avg}^{CDC}=\frac{IBL_{CPU}+IBL_{mem}+IBL_{net}}{N}
\end{equation}
(7). average running times:
Average running time of proceeding same amount of tasks can be compared for different scheduling algorithms.\\
(8). makespan:
In CloudSched, it is defined as the maximum load (or average utilization) on all PMs, and in some other simulators, it is defined as the longest processing time on all PMs.\\
(9). utilization efficiency: It is defined as (the minimum load on any PM) divides (maximum load on any PM) in this case.\\
\subsection{Metrics for Energy-efficiency}
\noindent (1). energy consumption model:\\
Most of energy consumption in data centers is from computation processing, disk storage, network, and cooling systems. In [5], authors proposed a power consumption model for blade server:
\begin{equation}
\small
14.5+0.2U_{CPU}+(4.5e^{-8})U_{mem}+0.003U_{disk}+(3.1e^{-8})U_{net}
\end{equation}
where $U_{CPU}, U_{mem}, U_{disk}, U_{net}$ are utilization of CPU, memory, hard disk and network interface respectively. From this formulation, it is observed that except CPU, the other factors such as memory, hard disk and network interface have very small impact on total energy consumption.

In [3], authors found that CPU utilization is typically proportional to the overall system load, hence proposed a power model as follows:
\begin{equation}
P(U)=kP_{max}+(1-k)P_{max} U
\end{equation}
where $P_{max}$ is the maximum power of a server; $k$ is the fraction of power when a server is idle, and studies show that on average the $k$ is about 0.7; and $U$ is the CPU utilization.
%According to the SPECpower benchmark, for the fourth quarter of 2010, the average energy consumption at 100% utilization for servers %consuming less than 1000 W was approximately 259 W.

In GreenCloud, Dynamic Voltage/Frequency Scaling (DVFS) is considered, the power consumption of an average server can be expressed as follows:
\begin{equation}
P=P_{fixed}+P_f \times f^3
\end{equation}
where $P_{fixed}$ accounts for the portion of the consumed power which does not scale with the operating frequency $f$, while $P_f$ is a frequency-dependent CPU power consumption.

The energy consumed by a switch and all its transceivers can be defined as:
\begin{equation}
\footnotesize
P_{switch} = P_{chassis} + n_{linecards}+P_{linecard}+\sum_{r=0}^{R} n_{ports,r}+ P_r
\end{equation}
where $P_{chassis}$ is related to the power consumed by the switch hardware, $P_{linecard}$ is the power consumed by any active network line card, $P_r$ corresponds to the power consumed by a port (transceiver) running at the rate $r$.

In real environment, the utilization of the CPU may change over time due to the workload variability. Thus, the CPU utilization is a function of time and is represented as u(t). Therefore, the total energy consumption by a physical machine ($E_i$) can be defined as an integral of the energy consumption function over a period of time as:
\begin{equation}
E_i=\int_{t_0}^{t_1} P(u(t))dt
\end{equation}
When the average utilization is adopted, u(t)=u, then $E_i$=$P(u) (t_1-t_0)$.\\
(2). The total energy consumption of a Cloud data center: The energy consumption is computed as the sum of energy consumed by all PMs:
\begin{equation}
E_{cdc}=\sum_{i=1}^{n} E_i
\end{equation}
It should be noted that the energy consumption of all VMs on PMs is included. \\
(3). The total number of PMs used: This is the total number of PMs used for the given set of VM requests. It is important for energy-efficiency.\\
(4) The total power-on time of all PMs used: According to the energy consumption equation of each PM, the total power-on time is a key factor. \\

\subsection{C/P (Cost/per task) Metric}
In iCanCloud, in order to deal with the complexity level added by an infrastructure following a pay-as-you-go basis, the C/P metric is defined as:
\begin{equation}
C/P=CT=\frac{C_hT_{exe}I}{iN_c^2} \lfloor{\frac{T_{exe}I}{iN_{vm}N_c} }\rfloor
\end{equation}
where $T_{exe}$ is the task execution time, the values of $I$ and $i$ correspond to the whole tracing interval and the tracing interval per task, that is, the grain of the application. On the other hand, $N_{vm}$ and $N_c$ are the number of Virtual Machines and number of cores per Virtual Machine, $C_h$ is the machine's usage price per hour. In this way, the best infrastructure setup would be that which produced the lowest C/P value.
\subsection {Confidence Interval}
Confidence intervals can be calculated for different metrics as follows: Let $x_1, x_2, x_3,..., x_n$ be the calculated metrics (such as $IBL_{tot}$ ~and~ $E_{cdc}$ values etc.) from $n$ times of repeated simulations. Then the mean is
\begin{equation}
x_{mean}=\frac{1}{n}\sum_{i=1}^{n} x_i
\end{equation}
and the standard deviation $s$ is
\begin{equation}
s=\sqrt{\frac{\sum_{i=1}^{n} (x_{mean}-x_i)^2}{n-1}}
\end{equation}
and the confidence interval at 95$\%$ confidence (normal distribution) is given by
\begin{equation}
(x_{mean}-1.96\frac{s}{\sqrt{n}}, x_{mean}+1.96\frac{s}{\sqrt{n}})
\end{equation}
\begin{table*}
\scriptsize
\caption{Metrics Comparison Guideline}
\begin{center}
\begin{tabular}{l|l|l}
\hline Metrics &Optimization Objectives &Simulators
\\\hline
\hline average resource utilization & maximizing resource utilization & All Four \\
\hline total number of PMs (hosts) need & maximizing resource utilization & All Four \\
\hline average CPU utilization & load balancing & All \\
\hline average utilization of all CPUs in a cloud datacenter & load balancing & All Four\\
\hline integrated load imbalance value of a server & load balancing & CloudSched\\
\hline imbalance value of all CPUs & load balancing &  CloudSched\\
\hline average imbalance value a physical server  & load balancing & CloudSched \\
\hline average imbalance value of a Cloud datacenter & load balancing & CloudSched\\
\hline total simulation time &All & All Four\\
\hline makespan or longest processing time & load balancing& CloudSim, CloudSched\\
\hline energy consumption model& energy-efficiency& CloudSim, GreenCloud, CloudSched\\
\hline total energy consumption of a Cloud data center & energy-efficiency & CloudSim, GreenCloud, CloudSched\\
\hline total number of PMs used	& energy-efficiency& CloudSim, GreenCloud, CloudSched\\
\hline total power-on time of all PMs & energy-efficiency & CloudSim, GreenCloud, CloudSched\\
\hline cost / per task & C / P	& iCanCloud\\
\hline confidence interval & confidence interval & CloudSched\\
\hline
\end{tabular} \\
\end{center}
\end{table*}

\section{Comparison 5: Performance Evaluation}
In this section, we will discuss the performance comparison of iCanCloud and CloudSim, CloudSched and CloudSched with a focus on the scalability. We also compare the typical outputs of all compared simulators.
\subsection{Performance Comparison of iCanCloud and Cloudsim}
\subsubsection{Experimental Environment Settings}
In the comparison between iCanCloud and CloudSim, jobs in CloudSim are modeled by configuring input size, processing length and output size. The jobs in the simulation experiments have 5 MB input size, 30MB output size, 1,200,000 MI processing length. In addition, jobs would take advantage of all the available CPU capacity on VMs and the VMs they used are 9,500 MIPS. Of course, a new application model is developed in iCanCloud to execute the same functionality as CloudSim. The experimental environment is on a computer with a CPU core i3 and 4GB of RAM memory.
\subsubsection{Performance Comparison}
Fig. 6(a) demonstrates the execution time comparison of CloudSim and iCanCloud, the x-axis presents the number of jobs executed in each experiment, y-axis presents the VMs number and its type, and z-axis presents the time required to execute each experiment (measured in seconds) in log-scale. It's obvious that both simulators need more execution time when increasing the number of jobs, while these simulators would have different impact when increasing the VMs number. When the VMs number is more than 2500, the execution time keeps stable in iCanCloud, while the execution time is influenced directly by both VMs number and jobs number. In most experimental cases with jobs amount less or equal to 50000, iCanCloud is faster than CloudSim, and in all tests with 250k jobs, iCanCloud is faster. Under all tests, iCanCloud shows better performance in execution time than CloudSim.

Fig. 6(b) presents the memory consumption comparison in each experiment for CloudSim and iCanCloud. It can be noticed in this graph that iCanCloud requires more memory than CloudSim. Up to 1000 VMs, the amount of memory required by both simulators is similar. However, when using more than 1000 VMs, the amount of memory required by iCanCloud goes up much faster than CloudSim.

In general, iCanCloud is faster in large scale experiments and provides better scalability, but requires more memory than CloudSim.
\begin{figure*} [htp!]
\begin{center}
{\includegraphics [width=1.0\textwidth,angle=-0] {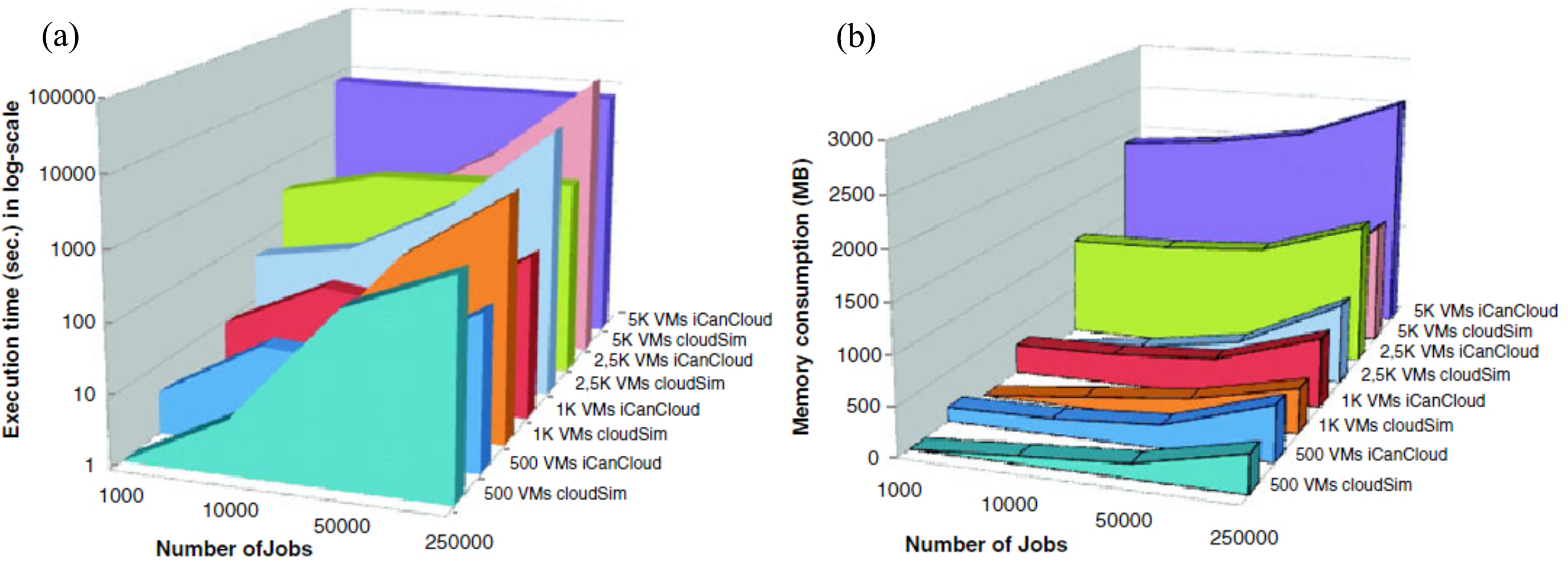}}
\caption{Performance comparison of CloudSim vs. iCanCloud \cite{IEEEhowto:Nunez}}
\end{center}
\end{figure*}

\subsection{Performance Comparison of CloudSim and CloudSched}
\subsubsection{Experimental Environment Settings}
In the comparison between CloudSim and CloudSched, the comparison is a bit complex than the comparison in section 7.1, a new construct method is created with start-time and end-time parameters, which refers to the lifecycle of a request. The file size of request represents the required capacity of all requests. The start-time, end-time generation approaches are same, servers (named VMs in CloudSim) and requests (named cloudlet in CloudSim) both adopt the EC2 specifications. List Scheduling algorithm is implemented in both simulators, in which requests would be allocated to a PM with the lowest utilization. The experimental environment is based on a Dell computer with a CPU core i5 and 8GB of RAM memory.
\begin{figure*} [htp!]
\begin{center}
{\includegraphics [width=1.0\textwidth,angle=-0] {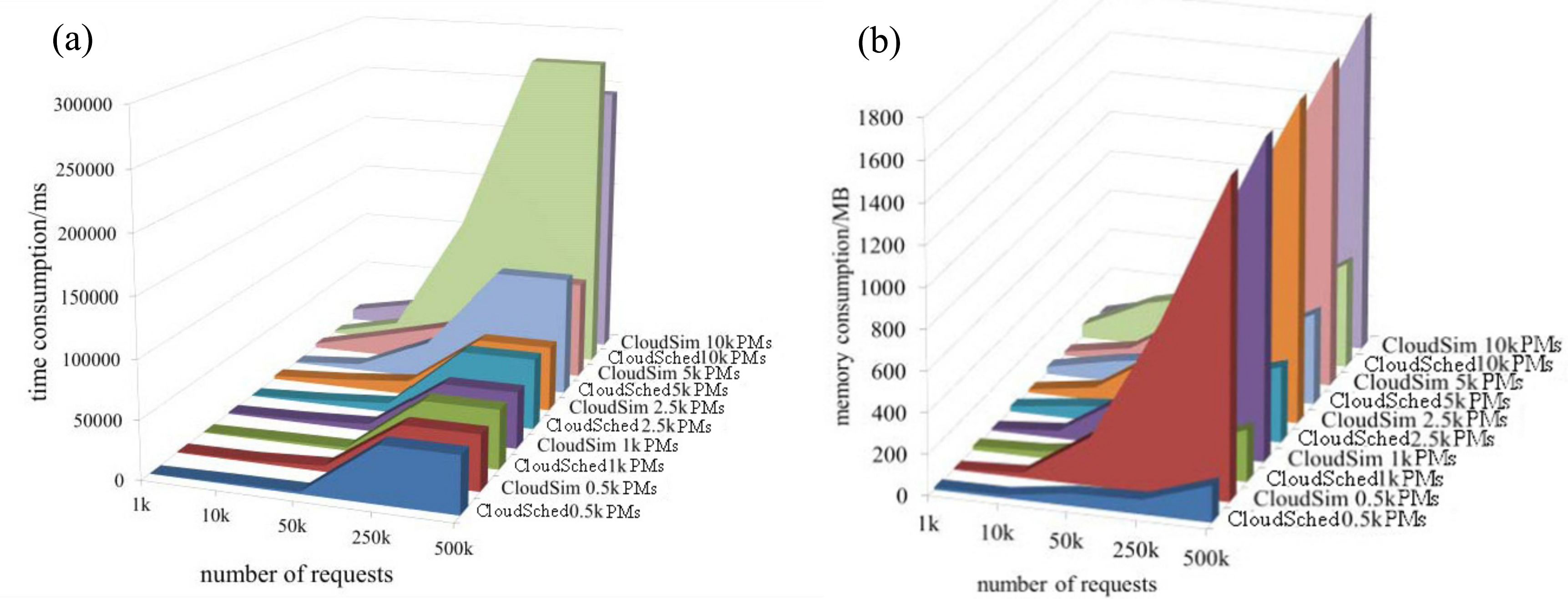}}
\caption{Performance Comparison of CloudSim vs. CloudSched}
\end{center}
\end{figure*}

\subsubsection{Performance Comparison}
Fig. 7(a) illustrates the time consumption of each experiment, where x-axis shows the requests number in each experiment for CloudSim and CloudSched, y-axis shows the number of PMs and simulators they belong to, and z-axis shows the time required in millisecond unit to simulate each experiment. It is also apparently observed that larger number of requests and number of PMs  need more time in both simulators. When the number of VMs is less than 10,000, CloudSched always costs less time to complete simulation. As for the numbers of VMs are 50,000 and 25,000, CloudSched takes less time than CloudSim, while CloudSched takes longer time when the number of PMs is more than 5,000. As the ratio of the number of VMs  to the number of PMs increases, like 500,000: 500, CloudSim shows its strength. Note that the ratio of VMs to PMs may be varying from a few to a few tens in a real cloud data center.

Fig. 7(b) shows the memory consumption comparison of each simulation in CloudSim and CloudSched. In cases when the VMs number is relative small, like from 1,000 to 10,000. CloudSched needs a little more memory, several megabytes, to execute simulations. While as the requests number becomes larger, CloudSched costs much less memory than CloudSim, the large difference happens when the request number is 500,000. The reason is that the VM and PM model in CloudSched is simpler than the models in CloudSim.

In general, CloudSched costs less time when the ratio of the number of VM requests  to the number of PMs  is not too large (like below 100) and costs much less memory than CloudSim.

\subsection {Typical Outputs Compared}
\begin{figure*} [htp!]
\begin{center}
{\includegraphics [width=1.0\textwidth,angle=-0] {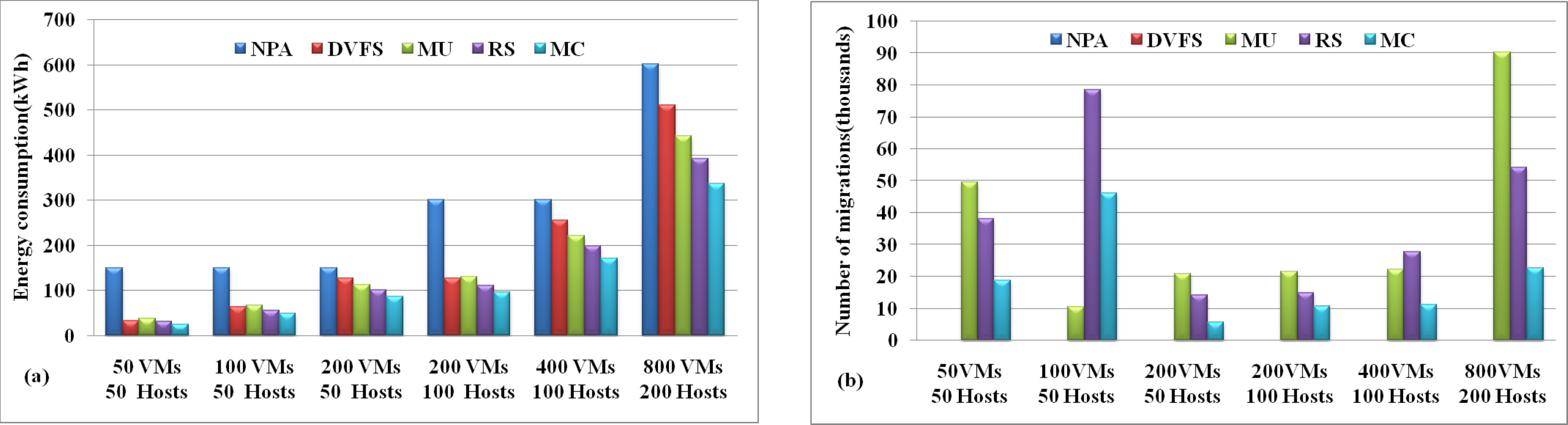}}
\caption{Typical Output of CloudSim}
\end{center}
\end{figure*}
In Fig. 8, we compare the performance of four energy-conscious resource management strategies against a benchmark technique NPA (NonPowerAware). In the benchmark technique, the processors can be operated at higher possible processing capacity as 100\% and do not consider energy-optimization during provisioning of VMs to hosts. The first energy-conscious strategy for comparison is DVFS enabled, which means that the VMs are resized during the simulation based on the dynamics of CPU utilization of the host. The other strategies are extensions of DVFS policy: MU (minimum utilization) strategy allocates VMs on the minimal utilization nodes; RS (Random selection) strategy randomly allocates VMs to hosts; MC (maximum correlation) strategy allocates VMs on the maximal correlation hosts. All these extensive strategies enable the idle nodes into sleep mode to save total energy and live migration of VMs every 5s for adapting to the allocation. VMs can be migrated to another host, if this operation will reduce energy consumption. In our simulation, the requests come randomly and we vary the number of hosts and VMs to obtain data for the energy consumption and the number of migrations. From our simulations,  MC strategy shows the best energy-efficient effects as shown in Fig. 8 (b). As for the number of migrations, NPA and DVFS both have no migrations, while  in other three strategies, MC strategy takes least number of  migration in most cases. The data  shown in Figure 8 is the average of  5 times of repeated simulation.

\begin{figure} [htp!]
\begin{center}
{\includegraphics [width=0.48\textwidth,angle=-0] {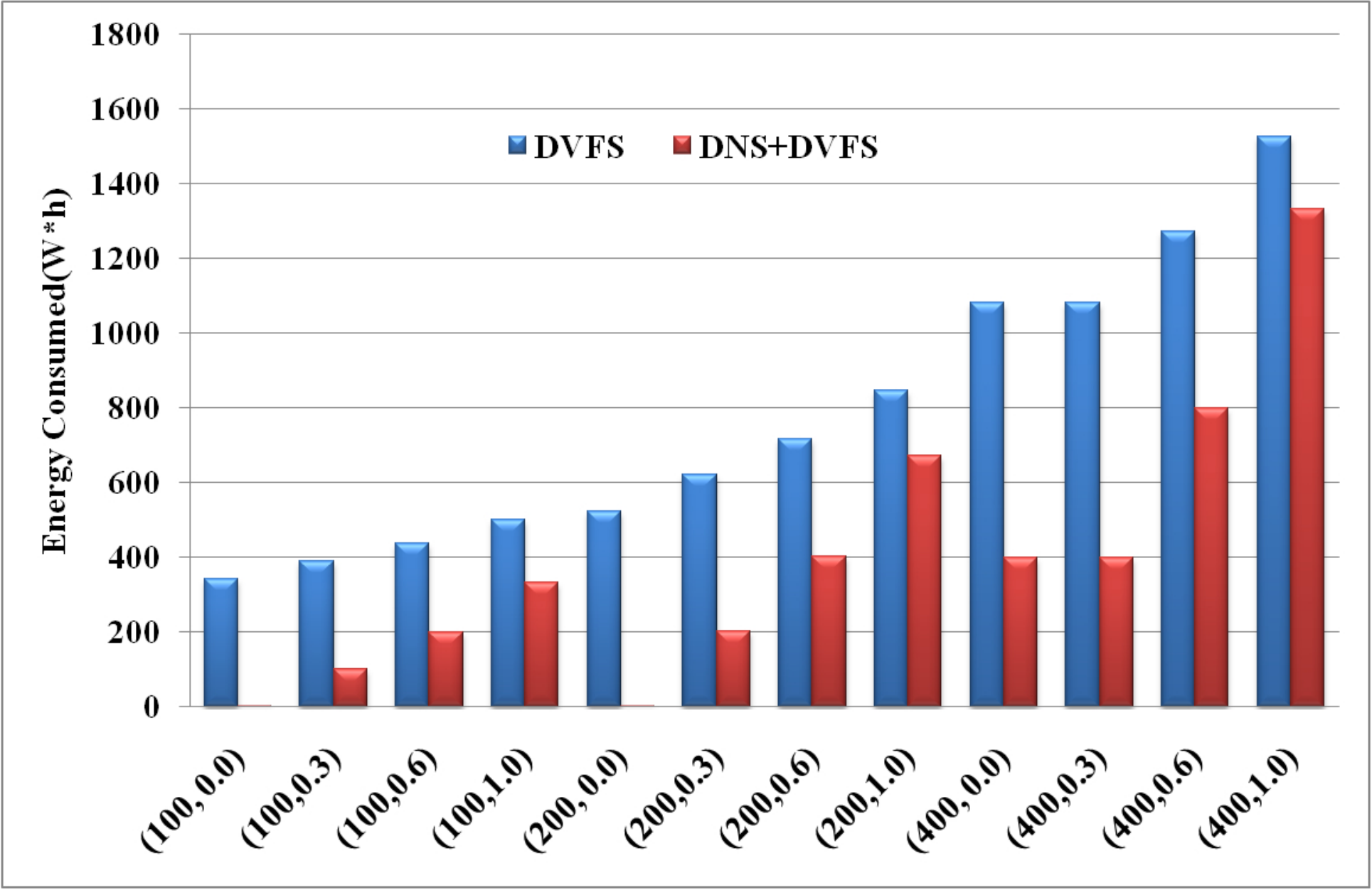}}
\caption{Typical Output of GreenCloud}
\end{center}
\end{figure}
In Fig.9 with GreenCloud simulations, we collect the total energy consumption under variable data center load (varying from 0.0, 0.3, 0.6 to 1.0) and variable number of  servers (varying from 100 to 400) both for DVFS only and DNS+DVFS power management schemes. The x-axis formats like (100, 0.0) represents the tests with 100 servers and 0.0 load, and (400, 1.0) shows tests with 400 servers and 1.0 load. In our simulations, we set the type of workloads as HPC (High Performance Computation) and the results gathered are averaged over 5 runs with the random number generator. From the bar chart, generally, it's obvious that the total energy consumption increases as the number of servers increases. It also demonstrates that the DVFS scheme shows itself little sensitive to the input load of servers, while by contrast the DNS+DVFS scheme shows precise sensitive to variable load. We also observe that under same number of servers and identical loads, the DNS+DVFS scheme saves more energy than DVFS scheme.

\begin{figure*} [htp!]
\begin{center}
{\includegraphics [width=1.0\textwidth,angle=-0] {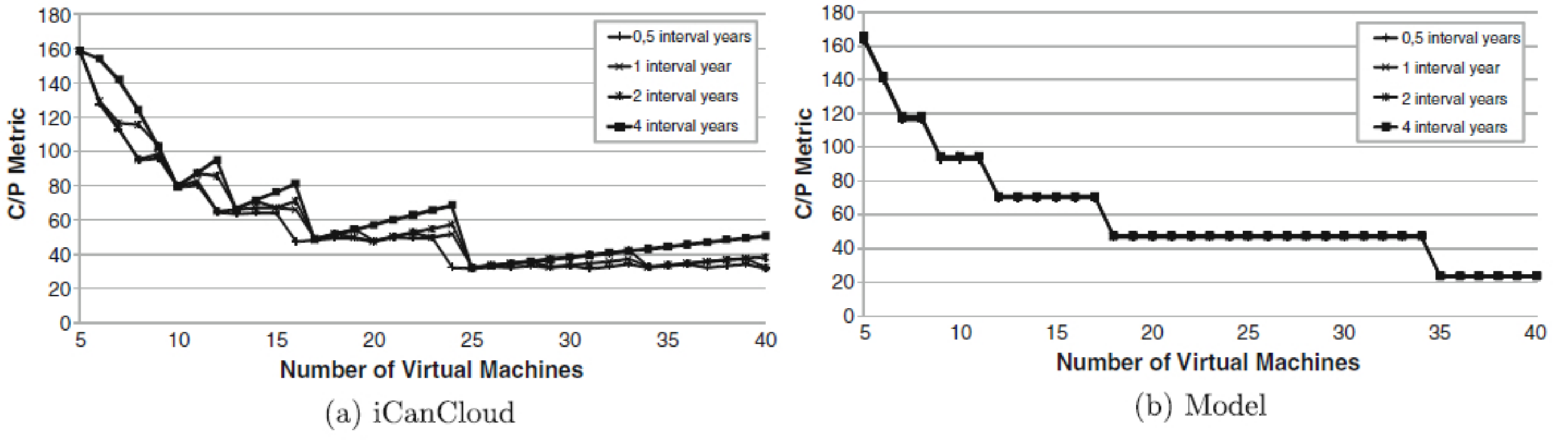}}
\caption{Typical Output of iCanCloud \cite{IEEEhowto:Nunez}}
\end{center}
\end{figure*}
In iCanCloud, Fig. 10 illustrates the results gathered by executing the model of Phobos application along with the results of the same application implemented on iCanCloud. The figure represents the C/P metric for the experiments, where the small instance type recommended by Amazon EC2 is provided, and the VMs number and tracing intervals are varied. From the results, we can notice that in some cases, using the same size for the interval (in years) and increasing the VMs number, causes an upward trend in the C/P metric. Then, increasing the VMs number provides the same execution time, which contributes to a increasing of the cost for this configuration. Besides that, the mathematical model does not represent the time spent on performing I/O operations. Because that there are still some problems for installation of current release of iCanCloud, we cannot test more data but using results in its original publication.

\begin{figure} [htp!]
\begin{center}
%\hfill
{\includegraphics [width=0.48\textwidth,angle=-0] {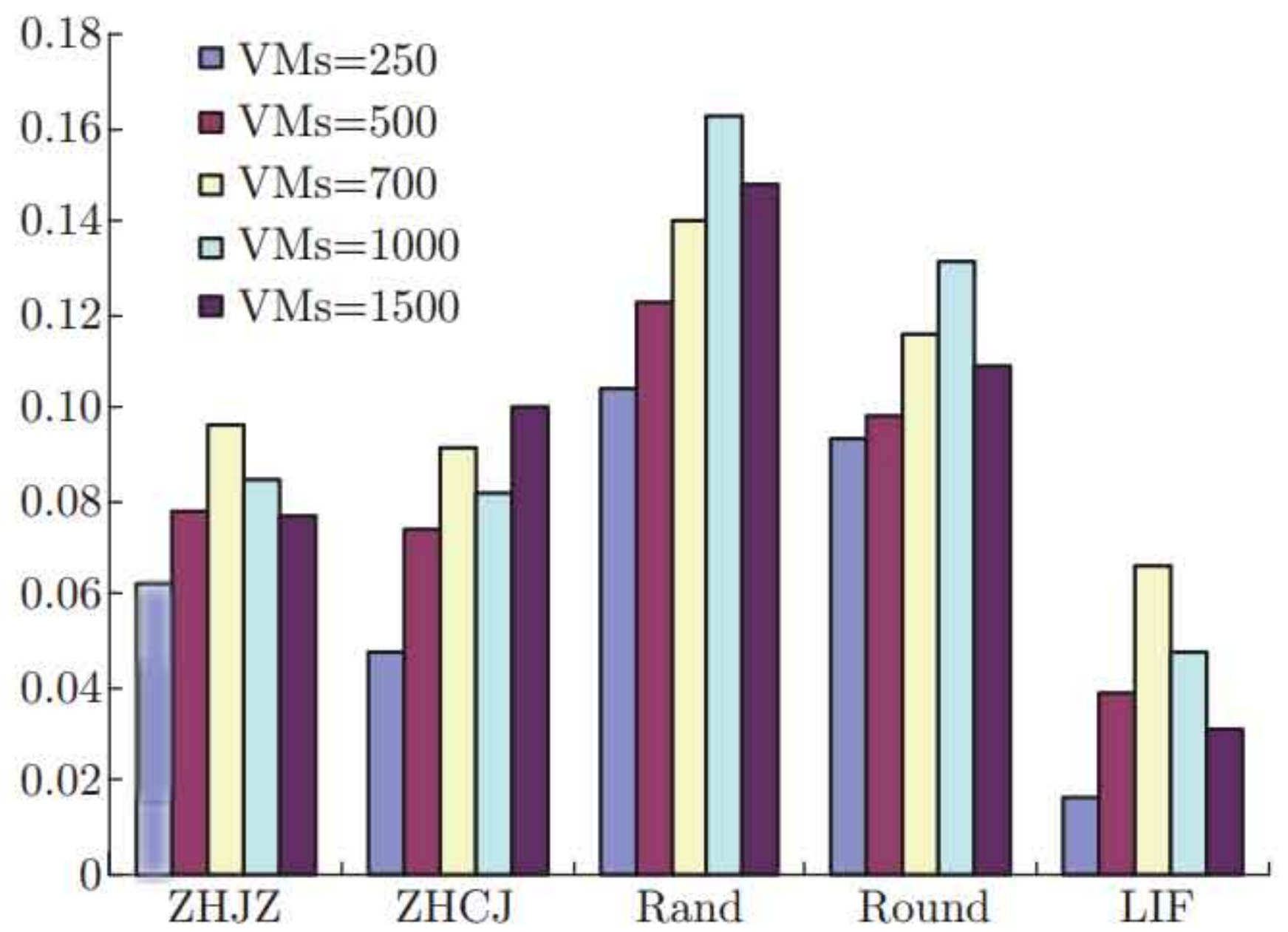}}
%\hspace*{\fill}
% \hfill \includegraphics [width=5\textwidth,angle=-0]{RunningTime.eps}\hspace*{\fill}
\caption{Typical CloudSched Outputs: Average Imbalance Values of a Cloud Data Center when PMs=100}
\end{center}
\end{figure}
In CloudSched, Fig. 11 shows average imbalance level of a cloud data center and five different scheduling algorithms for load balancing are compared. ZHCJ algorithm introduced in \cite{IEEEhowto:Wood}, ZHJZ algorithm \cite{IEEEhowto:Zheng}, LIF algorithm \cite{IEEEhowto:Tian2013}, Rand algorithm, and Round-Robin (Round) are compared. In these simulations, different requests are generated as follows: the total numbers of arrivals (requests) can be randomly set; all requests follow Poisson arrival process and have exponential length distribution; the maximum length of requests can be set; for each set of inputs (requests), simulations are run six times and all the results shown in this paper are the average of the six runs. In these simulations, the number of PMs  is fixed as 100, the number of requests is varying from 250 to 1500, and a PC with 2 GHz CPU, 2 GB memory is used for all simulations. From these simulations, we observe that LIF algorithm outwits other four algorithms with average imbalance values, which shows that LIF has a better load balance effects than others.

%The input data of user requests is generated using program by considering equal probabilities of above mentioned eight types of VMs. Of %course, different (random) probabilities of different types of VMs can be generated. For steady-state analysis, a warm-up period (initial %2000 requests) is used to drop the transient period. It can be seen that LIF algorithm has lowest average imbalance level when the total %VMs number and PMs are varying. Similar outputs for other load balancing and energy saving cases can be obtained as well.
\section{Conclusions and Future Work}
In this paper, we mainly compare four open source simulators, namely CloudSim, GreenCloud, iCanCloud and CloudSched. These simulators can simulate the cloud data center scenarios from different layers in the cloud computing architecture. From their architectures, elements modeling, simulation process, performance metrics and outputs, we provide detailed comparisons about these simulators. Considering the complexity of networks and the difficult to control the network traffics, simulators are crucial tools for research. We can see that none of them is perfect for all aspects and there are still much work to do to improve. One suggestion is to use different tools or their combinations for different optimization objectives such as load balance and energy-efficiency.  For future work, there are still quite a few challenging issues for cloud simulating:
\begin{itemize}
%\item
%\textbf{Generality and transparency}. The simulator should be platform-independent and can be applied (transplanted) to any platform.
\item
\textbf{Modeling different Cloud layers}. As we compared in the paper, each tool may focus on one layer. Currently there is still lack of tools that can model all Cloud layers (IaaS, PaaS and SaaS).
\item
\textbf{High extensibility}. When new policies and algorithms are added, modular design of the simulators can assure that new modules can be easily added, currently the four simulators still need improving this.
\item
\textbf{Easy to use and repeatable}. The simulators should enable users to set up simulation easily and quickly with easy to use graphical user interfaces and outputs. It can accept inputs from text files and output to text files; can save simulation inputs and outputs so that modelers can repeat experiments, ensuring that repeated simulation yield identical results.
\item
\textbf{Considering user priority}. This is a  real requirement. Currently the four simulators do not consider this yet. Different priority policies can be created for users to have different priorities for certain types of VMs, so that more realistic scenarios  can be considered.
\item
\textbf{Supporting multiple or federated data centers}. The simulator should be able to reflect and model the multiple or federated data centers in real world. CloudAnalyst provides a framework by extending CloudSim and there is still much work to improve.
\end{itemize}

\section*{Acknowledgment}
This research is sponsored by the National Natural Science Foundation of China (NSFC) (Grant Number:61150110486).
\end{document}